\documentclass{article}
\usepackage{ijcai22}

\usepackage{times}
\usepackage{soul}
\usepackage{url}
\usepackage[hidelinks]{hyperref}
\usepackage[utf8]{inputenc}
\usepackage[small]{caption}
\usepackage{graphicx}
\usepackage{amsmath}
\usepackage{amsthm}
\usepackage{booktabs}
\usepackage{algorithm}
\usepackage{algorithmic}
\usepackage{subfigure}
\usepackage{graphicx}

\title{Unbiased Implicit Feedback via Bi-level Optimization}

\author{
Can Chen$^1$
\and
Chen Ma$^2$
\and
Xi Chen$^{3}$
\and
Sirui Song$^1$
\and
Hao Liu$^4$
\And
Xue Liu$^1$
\affiliations
$^1$ McGill University
$^2$ City University of Hong Kong
$^3$ Huawei Noah's Ark Lab\\
$^4$ The Hong Kong University of Science and Technology~(Guangzhou)
\emails
\{can.chen, sirui.song\}@mail.mcgill.ca,
chenma@cityu.edu.hk,
xi.chen11@mcgill.ca,\\
liuh@ust.hk,
xueliu@cs.mcgill.ca
}
\begin{document}

\maketitle

\begin{abstract}
Implicit feedback is widely leveraged in recommender systems since it is easy to collect and provides weak supervision signals.
Recent works reveal a huge gap between the implicit feedback and user-item relevance due to the fact that implicit feedback is also closely related to the item exposure.
To bridge this gap, existing approaches explicitly model the exposure and propose unbiased estimators to improve the relevance.
Unfortunately, these unbiased estimators suffer from the high gradient variance, especially for long-tail items, leading to inaccurate gradient updates and degraded model performance.
To tackle this challenge, we propose a low-variance unbiased estimator from a probabilistic perspective, which effectively bounds the variance of the gradient.
Unlike previous works which either estimate the exposure via heuristic-based strategies or use a large biased training set,
we propose to estimate the exposure via an unbiased small-scale validation set.
Specifically, we first parameterize the user-item exposure by
incorporating both user and item information, and then construct an unbiased validation set from the biased training set.
%i 
By leveraging the unbiased validation set, we adopt bi-level optimization to automatically update exposure-related parameters along with recommendation model parameters during the learning.
Experiments on two real-world datasets and two semi-synthetic datasets verify the effectiveness of our method.
%
%Our code is available at \url{https://anonymous.4open.science/r/CODE-50C2/}.

\end{abstract}

% \begin{small}
% \begin{equation}
% \begin{aligned}
% \mathcal{L}(\omega) &= -\sum_{(u,i)\in \mathcal{D}}{\tilde{R}_{ui}\log p(\tilde{R}_{ui}=1|\omega)
% \\ &
% + (1-\tilde{R}_{ui})\log p(\tilde{R}_{ui}=0|\omega)}
% \end{aligned}
% \end{equation}
% \end{small}

% \begin{equation}
%     \begin{aligned}
%     \mathcal{L}(\omega) &= -\sum_{(u,i)\in \mathcal{D}}{\tilde{R}_{ui}\log p(\tilde{R}_{ui}=1|\omega)
% \\ &
% + (1-\tilde{R}_{ui})\log p(\tilde{R}_{ui}=0|\omega)}
%     \end{aligned}
% \end{equation}
%\textcolor{blue}{[MA: maybe change the title to Noise-free Top-K recommender via Bi-level Optimization.]}

\section{INTRODUCTION}
Recent years have witnessed the fast development of the recommender system.
It has been successfully deployed in many web services like E-commerce and social media. 
Learning from historical interactions, a recommender system can predict the relevance or preference between users and items. 
Based on which, the system recommends items that the user may prefer.
To enable these, there are two types of feedback: explicit feedback and implicit feedback.
Explicit feedback can be the ratings on items that explicitly represent the preferences of the users.
However, the collection of explicit feedback requires the user active participation, which makes explicit feedback unavailable in most real-world scenarios.
Compared with explicit feedback, implicit feedback such as clicks is widely used because of its ubiquity and wide availability.
Though easier to collect, implicit feedback is one-sided and positive only~\cite{yang2018unbiased}, which means the recommender can only observe the users' interactions with relevant items.
A missing link between a user and an item can either be that the user dislikes the item or that the item is not exposed to the user~\cite{liang2016modeling}.
%A missing item may mean the user dislikes the item or the item is not exposed to the user.

%
Many important works have tried to improve recommendation performances in implicit feedback by explicitly modeling the user-item exposure.
For example,  \cite{yang2018unbiased} find implicit feedback subject to popularity bias, and propose an unbiased evaluator based on the Inverse-Propensity-Scoring (IPS) technique  \cite{joachims2016counterfactual}.
which significantly reduces the evaluation bias.
Exposure matrix factorization (ExpoMF)  \cite{liang2016modeling} introduce exposure variables to build a probabilistic model, and
consider external information when estimating exposure.
%
%ExpoMF separately captures the user-item exposure and relevance and achieves significant improvement.
%
Yet,  \cite{saito2020unbiased} find that ExpoMF is biased towards popular items and yields unsatisfied results for rare items. 
Based on IPS, \cite{saito2020unbiased} propose an unbiased estimator to maximize the user-item relevance.
In \cite{saito2020unbiased}, both the user-item relevance and the user-item
exposure are modeled as Bernoulli random variables, and the click probability is the product of exposure probability and relevance probability.
\cite{saito2020unbiased} better achieve the objective of the unbiased recommendations than alternatives  \cite{liang2016modeling,hu2008collaborative}.
Using the same unbiased estimator in \cite{saito2020unbiased}, \cite{zhu2020unbiased} propose a combinational joint learning framework to more accurately estimate exposure.
%for better exposure estimation.

%
However, we find \cite{saito2020unbiased}\cite{zhu2020unbiased} suffers from the high gradient variance problem.
Inaccurate gradient updates occur in the learning process, which degrades the model performance.
%
%Expo-MF proposes to model user-item exposure as a latent variable and learn it with the help of external information. However, Expo-MF is biased with regard to rare items.
%
%Besides, previous works  \cite{saito2020unbiased,zhu2020unbiased} usually estimated the user-item exposure only using the item information, but the estimation should also depend on the user since active users should have larger exposure.
%
Moreover, these existing approaches \cite{hu2008collaborative}\cite{yang2018unbiased}\cite{saito2020unbiased} \cite{zhu2020unbiased} adopt some simple heuristic-based strategies or only leverage the biased training set to estimate exposure, which inevitably leads to a biased recommendation model.

To tackle the high gradient variance problem, we develop a low-variance estimator from a probabilistic perspective.
To better estimate exposure, we model exposure by incorporating both user and item information and construct a small amount of unbiased validation set to guide exposure estimation.
% exposure parameters automatically.
%
Specifically, with the unbiased set, we introduce bi-level optimization  \cite{colson2007overview} with exposure parameters as the outer variable and relevance parameters as the inner variable, to update exposure parameters automatically.
Overall, we propose \textbf{UBO} (\textbf{U}nbiased Implicit Feedback via \textbf{B}i-level \textbf{O}ptimization) to update exposure parameters simultaneously with relevance parameters.
We further analyze the inner mechanism of bi-level optimization in UBO and compare bi-level optimization with other optimization methods.
We verify the effectiveness of UBO on both real-world and semi-synthetic datasets.

To summarize, our work has three contributions:
\begin{enumerate}
    \item We propose a low-variance unbiased estimator which effectively bound gradient variance.
    %and connect exposure estimation not only with the item information but also with the user information. 
    \item We connect exposure estimation to both user and item information and introduce bi-level optimization to update exposure parameters by leveraging an unbiased set.
    %\item Guided by a small unbiased validation set, we introduce bi-level optimization to update exposure simultaneously with relevance. 
    \item Furthermore, we give a natural interpretation of why bi-level optimization works by gradient analysis, and compare it with other optimization methods to better understand its necessity.
    %\item Experiments on two real-world datasets and two semi-synthetic datasets demonstrate the effectiveness of UBO. %\textcolor{red}{XC: not big enough as a contribution.}
\end{enumerate}

% The structure of the paper is as follows.
% %
% In Section~\ref{method}, we introduce some notations and illustrate the UBO algorithm.
% %
% Then we analyze the experimental results on two real-world datasets in Section~\ref{real} and two semi-synthetic datasets in Section~\ref{semi}.
% % In Section~\ref{real}, we conduct experiments on two real-world datasets to answer two research questions.
% % %
% % In Section~\ref{semi}, we conduct experiments on two semi-synthetic datasets to investigate whether UBO learns exposure correctly.
% %
% In Section~\ref{related}, we discuss related work from two perspectives: debiasing and bi-level optimization in the recommender system.
% %
% In Section~\ref{conclusion}, we summarize our findings.
% previous related work and the relationship with our method. In
% Section 3 we present the CausE approach. In Section 4 we present
% the experimental setup and the results on the MovieLens dataset.
% In Section 5, we summarize our findings and conclude with future
% directions of research.

\section{METHOD}
\label{method}
%In this section, we introduce the development of our algorithm and details on its design. 
In this section, we begin by introducing some preliminaries including notations and the previous unbiased estimator. 
Then we show the high gradient variance problem and derive our low-variance unbiased estimator from a probabilistic perspective. 
Further, we parameterize the user-item exposure by considering both user and item information and propose to use a small unbiased validation set to guide exposure estimation via bi-level optimization.
%
%
%the modeling of the user-item exposure $m_{ui}$

\subsection{Preliminaries}
\noindent \textbf{Notations.} Assume we have an implicit feedback dataset $\mathcal{D}$ with $N$ users indexed by $u$ and $M$ items indexed by $i$. 
Let $\tilde{R}_{ui}$ denote the observed feedback between $u$ and $i$.
$\tilde{R}_{ui}=1$ indicates positive feedback, while $\tilde{R}_{ui}=0$ indicates either positive unlabeled feedback or negative feedback.

To precisely formulate implicit feedback, \cite{saito2020unbiased} introduces two kinds of Bernoulli random variables $R_{ui}$ and $O_{ui}$.
${R}_{ui}$ represents the user-item relevance between $u$ and $i$ with $\gamma_{ui}$ as the Bernoulli parameter.
${R}_{ui}=1$ means $u$ and $i$ are relevant, and $R_{ui}=0$ means $u$ and $i$ are not relevant.
Similarily, ${O}_{ui}$ represents the user-item exposure between $u$ and $i$ with $m_{ui}$ as the Bernoulli parameter.
${O}_{ui}=1$ means $i$ is exposed to $u$, and vice versa.
We denote $\bar{m}_{ui}$ as the estimated exposure
between $u$ and $i$ in the following paper.
Note that both $R_{ui}$ and ${O}_{ui}$ can not be observed in implicit feedback.
$\tilde{R}_{ui}$ is also a Bernoulli random variable:
\begin{equation}
\label{eq1}
    \tilde{R}_{ui} = R_{ui}O_{ui} 
\end{equation}
The Bernoulli parameter of $\tilde{R}_{ui}$ can be written as:
\begin{equation}
P(\tilde{R}_{ui}=1) = m_{ui}\gamma_{ui}
\end{equation}
From Eq~(\ref{eq1}), we can see that a positive feedback $R_{ui}=1$ means that $i$ is exposed to $u$ and $u$ likes $i$.

The task of the implicit recommendation system is to provide an ordered
set of items for users based on the predicted user-item relevance. 
We use $p_{ui} = p({R}_{ui}=1|\omega)$ to represent the predicted user-item relevance where the relevance parameters $\omega$ include the user embedding $\omega_u$ and the item embedding $\omega_i$.
Since matrix factorization is the most widely used technique~\cite{koren2009matrix} in the recommender system, in this paper we compute the predicted user-item relevance as:
\begin{equation}
    p_{ui} = \sigma(\omega_u^T\omega_i)
\end{equation}
where $\sigma(\cdot)$ represents the sigmoid function.
Note that UBO can also be easily applied on other neural network based models~\cite{he2017neural}\cite{wang2019neural}.
% %
% Implicit feedback prediction as a binary classification task, and use the most widely used loss for classification: cross-entropy loss.
% %
% Note the following analysis still holds for other loss functions like the mean square loss.
% %
% %Yet, the most widely used method BPR is not compatible with this formulation due to its pairwise loss function.
% %
% Specifically, the cross-entropy loss can be written as:
% \begin{equation}
% \begin{aligned}
% \mathcal{L}(\omega) &= -\sum{\tilde{R}_{ui}\log p(\tilde{R}_{ui}=1|\omega)+\\& (1-\tilde{R}_{ui})\log  p(\tilde{R}_{ui}=0|\omega)}
% \end{aligned}
% \end{equation}
% %
% By minimizing $\mathcal{L}(\omega)$, $R_{ui}$ can be approximated by $\bar{R}_{ui}$.
% The task of implicit recommendation system is to provide an ordered set of items for users based on $\bar{R}_{ui}$.

%

\noindent \textbf{Unbiased Estimator.}  \cite{saito2020unbiased} find the top-k recommendation metrics such as the mean average precision \cite{yang2018unbiased} can not directly signify relevance, and thus are not proper to measure recommendation results.
To optimize the performance metric of relevance,  \cite{saito2020unbiased} propose an unbiased estimator from the IPS technique, and the log loss form can be written as:
%\vspace{-5pt}
\begin{small}
\begin{equation}
\begin{aligned}
\mathcal{L}_1(\omega)\!&=\!-\!\sum_{(u,i)\in \mathcal{D}}{\frac{\tilde{R}_{ui}}{\bar{m}_{ui}}\log p_{ui}\!
+\!(1-\frac{\tilde{R}_{ui}}{\bar{m}_{ui}})\log (1-p_{ui})}
\end{aligned}
\end{equation}
\end{small}
%\vspace{-5pt}
%
%\cite{zhu2020unbiased} adopt the same unbiased estimator.
%
%In this paper, we follow the definition of RelMF.
%
Once we have the expectation of $\mathcal{L}_1(\omega)$, we will find the optimal solution for $p_{ui}$ is $\gamma_{ui}$ given an accurate exposure estimation $\bar{m}_{ui} = m_{ui}$.
This proves this estimator unbiased.
In this paper, we mainly consider the log loss form since it is the most widely used form. 
Other loss forms such as the mean squared loss can be analyzed similarly.

\subsection{Proposed Unbiased Estimator}
\noindent \textbf{High gradient variance}.
The gradient of $\mathcal{L}_1(\omega)$ to $p_{ui}$ is:
\begin{alignat}{2}
\frac{\partial{\mathcal{L}_1(\omega)}}{\partial{p_{ui}}} = -[\frac{\tilde{R}_{ui}}{\bar{m}_{ui}}(\frac{1}{p_{ui}}+\frac{1}{1-p_{ui}})-\frac{1}{1-p_{ui}}]
\end{alignat}
The variance of $\frac{\partial{\mathcal{L}_1(\omega)}}{\partial{p_{ui}}}$ can be calculated by:
\vspace{-5pt}
\begin{equation}
    \begin{aligned}
    V(\frac{\partial{\mathcal{L}_1(\omega)}}{\partial{p_{ui}}})
    &= \frac{\gamma_{ui}(1-\bar{m}_{ui}\gamma_{ui})}{\bar{m}_{ui}p_{ui}^2(1-p_{ui})^2} \\
    \end{aligned}
\end{equation}
For rare items, $\bar{m}_{ui}$ can be very small so that $V(\frac{\partial{\mathcal{L}_1(\omega)}}{\partial{p_{ui}}})$ becomes unbounded.
This problem leads to inaccurate gradient updates and decreases the model performance.

\noindent \textbf{Low-variance unbiased estimator}. Instead of deriving from the IPS technique, which leads to the high gradient variance problem, we propose a low-variance unbiased estimator from a probalistic view. To be specific, we first write the cross-entropy loss as:
\vspace{-5pt}
% \begin{small}
% \begin{equation}
% \begin{aligned}
% \mathcal{L}(\omega) &= -\sum_{(u,i)\in \mathcal{D}}{\tilde{R}_{ui}\log p(\tilde{R}_{ui}=1|\omega)\\&+ (1-\tilde{R}_{ui})\log p(\tilde{R}_{ui}=0|\omega)}
% \end{aligned}
% \end{equation}
% \end{small}
\begin{small}
\begin{equation}
\begin{aligned}
\mathcal{L}(\omega) &= -\sum_{(u,i)\in \mathcal{D}}{\tilde{R}_{ui}\log p(\tilde{R}_{ui}=1|\omega)}\\ & +{ (1-\tilde{R}_{ui})\log p(\tilde{R}_{ui}=0|\omega)}
\end{aligned}
\end{equation}
\end{small}
Recall that we are caring about the user-item relevance prediction $p_{ui}$. 
From the probabilistic perspective, we have:\\
\vspace{-5pt}
\begin{equation}
\begin{aligned}
    p(\tilde{R}_{ui}=1|\omega) &=  {\bar{m}_{ui}p_{ui}}
\end{aligned}
\end{equation}
\begin{equation}
\begin{aligned}
    p(\tilde{R}_{ui}=0|\omega) &= {1-\bar{m}_{ui}p_{ui}}
\end{aligned}
\end{equation}
Our estimator is defined as:
\begin{small}
\begin{equation}\label{estimator}
\begin{aligned}
\mathcal{L}_2(\omega) & = -\sum_{(u,i)\in\mathcal{D}}
{\tilde{R}_{ui}{\log}(\bar{m}{_{ui}}{p_{ui}})} \\& + {(1-\tilde{R}{_{ui}}){\log} (1-\bar{m}_{ui} p_{ui})}
\end{aligned}
\end{equation}
\end{small}

After computing the expectation of $\mathcal{L}_2(\omega)$, we can easily find the optimal solution for $p_{ui}$ is also $\gamma_{ui}$ given an accurate exposure estimation $\bar{m}_{ui}=m_{ui}$.
This proves our estimator unbiased~(See Appendix A for details).
%.(See Appendix A for details), which 
%
Besides, our unbiased estimator yields better gradient properties for rare items.
%
%\noindent Specifically, we calculate the gradient as:
% \begin{alignat}{2}
% \frac{\partial{\mathcal{L}_2(\omega)}}{\partial{p_{ui}}} = -(\frac{\tilde{R}_{ui}}{p_{ui}}+\frac{(\tilde{R}_{ui}-1)\bar{m}_{ui}}{1-\bar{m}_{ui}p_{ui}})
% \end{alignat}
%
\noindent The variance of $\frac{\partial{\mathcal{L}_2(\omega)}}{\partial{p_{ui}}}$ is calculated as:
\begin{equation}
    \begin{aligned}
    V(\frac{\partial{\mathcal{L}_2(\omega)}}{\partial{p_{ui}}})
    &= \frac{\bar{m}_{ui}\gamma_{ui}(1-\bar{m}_{ui}\gamma_{ui})}{p_{ui}^2(1-\bar{m}_{ui}p_{ui})^2}
    \end{aligned}
\end{equation}
%
% \begin{alignat}{2}
% V(\frac{\partial{\mathcal{L}_2(\omega)}}{\partial{p_{ui}}}) = \frac{\bar{m}_{ui}\gamma_{ui}(1-\bar{m}_{ui}\gamma_{ui})}{p_{ui}^2(1-\bar{m}_{ui}p_{ui})^2}
% \end{alignat}
$V(\frac{\partial{\mathcal{L}_2(\omega)}}{\partial{p_{ui}}})$ stays bounded as $\bar{m}_{ui}$ becomes small, and thus this estimator yileds stable gradient updates.
We don't consider the possible high gradient variance problem on $p_{ui}=0$ or $p_{ui}=1$ since this occurs in both estimators.
Our estimator only solves high gradient variance problems related to $\bar{m}_{ui}$.
%
%Some clipping techniques \cite{schnabel2016recommendations} may be applied to $p_{ui}$ to reduce variance at the cost of introducing bias.
%

\subsection{Exposure Estimation}

\subsubsection{Exposure Modeling}
%\noindent \textbf{Exposure Modeling.}
%
%Using a $M\times N$ matrix to store all $O_{ui}$ entries is space-consuming so we need to design simpler modeling of $m_{ui}$.
%
It is not realistic to assign every $O_{ui}$ entry a learnable parameter to represent the user-item exposure due to the space limit, so we need a distributed representation for all $O_{ui}$ entries.
%$M\times N$ learnable matrxi to represent all $O_{ui}$ entries due to the space limit.
%
In this paper, we parameterize the user-item exposure $m_{ui}$ with one  MLP(multi-layer perceptron) and $N$ user-wise embeddings and connect exposure estimation with both user and item information. 
%which considers both user and item information.
%
On one hand, $m_{ui}$ is large if the item is popular, which means we should consider the item popularity when estimating $m_{ui}$.  
Note that the popularity~\cite{he2016fast} of the item $i$ can be approximated as:
\begin{equation}
\begin{aligned}
    \theta_{i} = (\frac{\sum_u{\tilde{R}_{ui}}}{max_{i}\sum_u{\tilde{R}_{ui}}})^{0.5}
\end{aligned}
\end{equation}
On the other hand, $m_{ui}$ becomes large if the item is exposed to the user often or the user is active, which means we should also consider the impact of the user.
We introduce a new user-wise embedding $e_u$ and use $\sigma(e_u^T\omega_i)$ to represent the user's impact.
% \begin{equation}
% \begin{aligned}
% u_i = \sigma(e_u^T\omega_i)
% \end{aligned}
% \end{equation}
%
Note that the introduced user embedding $e_u$ can be learned directly through external user information~\cite{liang2016modeling}, whereas, in this paper, we assume we do not have the external information, which is more general. 
To sum up:
\begin{equation}
\begin{aligned}
    m_{\alpha}(u, i, \omega) = r(\omega_{i})\sigma(e_{u}^T\omega_{i}) +(1-r(\omega_{i}))\theta_i
\end{aligned}
\end{equation} 
Here $r(\cdot)$ learns the trade-off between the impact of the user and the popularity of the item, and we use one layer MLP followed by a sigmoid function to parameterize $r(\cdot)$.
% is a one-layer MLP followed by a $\sigma$ and learns the trade-off between the user's impact and the popularity of the item. 
We use $\alpha$ to denote exposure parameters, which include the introduced user embeddings and the MLP parameters in $r$.
For convenience, we still use $\bar{m}_{ui}$ instead of $m_{\alpha}(u, i, \omega)$ to represent the estimated exposure in the following paper.

\subsubsection{Bi-level Optimization}
%\\
% \begin{algorithm}[tb]
% \caption{Bi-level Optimization Framework}
% \label{alg:algorithm}
% \textbf{Input}: the training set $\mathcal{D}_{train}$; the unbiased validation set $\mathcal{D}_{val}$; the max iteration $T$; other hyperparameters.   \\
% %\textbf{Parameter}: $\omega$: the relevance parameters(the user and the item embeddings); $\alpha$: the exposure parameters to parameterize $\bar{m}_{ui}$.\\
% \textbf{Output}: $\omega^*$
% \begin{algorithmic}[1] %[1] enables line numbers
% \FOR{$t \in range(0, T)$}  
% \STATE SampleMiniBatch $\mathcal{B}_{train}$from $\mathcal{D}_{train}$; 
% \STATE Update $\omega$ with $\mathcal{B}_{train}$ by Eq.(\ref{inner});
% \STATE Update $\alpha$ with $\mathcal{D}_{val}$ by Eq.(\ref{outer});
% \ENDFOR
% %\STATE \textbf{return} $\omega^{*}$
% \end{algorithmic}
% \end{algorithm}
%
% Previous works \cite{saito2020unbiased,schnabel2016recommendations} estimate the user-item exposure from the biased training set such as using the item popularity or the logistic regression.
% %
% Yet, the feedback $\tilde{R}_{ui}=0$ in the biased training set indicates either that $i$ is exposed to $u$ but $u$ dislikes $i$ or that $i$ is not exposed to $u$.
% %
% This introduces bias into the user-item exposure estimation.
%
%
%\noindent  
% Previous approaches only use the biased training set to estimate exposure, such as using the logistic matrix factorization or the item popularity.
% %
% Whether the item is exposed to the user is unknown in the unobserved feedback, which leads to inaccurate exposure estimation \cite{zhu2020unbiased}. 
% %
Previous work~\cite{hu2008collaborative,yang2018unbiased,saito2020unbiased,zhu2020unbiased} adopt some simple heuristics or only use the biased training set to estimate exposure, which inevitably results in a biased model.
%
%We first propose the exposure estimation should consider both user and item information and design modeling to represent this trade-off.
%
%Then we construct an unbiased validation set from the biased training set, with which we automatically update exposure parameters via bi-level optimization.
We propose to leverage a small unbiased validation set to guide exposure estimation via bi-level optimization. 
Specifically, we select the most popular positive item and negative item for each active user to form the unbiased validation set. 
The reason why the validation set can be treated as unbiased is that these items are very likely to be exposed to these users and we approximate $m_{ui}$ as 1 in the validation set. 
%
%We update $\alpha$ automatically by minimizing the validation loss $L_{val}(\omega^*(\alpha))$.
%

\noindent\textbf{Formulation}.
We use the proposed estimator Eq~(\ref{estimator}) to calculate the training loss $\mathcal{L}_{train}$ and the validation loss $\mathcal{L}_{val}$.
%
%
%Denote by $\mathcal{L}_{train}$ and $\mathcal{L}_{val}$ the training and the validation loss, respectively.
%
Given an unbiased training set, we obtain the optimal user and item embeddings $\omega^*$ by minimizing $\mathcal{L}_{train}(\omega)$.
Whereas in a biased training set, different user-item pairs have different exposure.
Thus for a biased training set, we need to first estimate the user-item exposure $\bar{m}_{ui}$ parametrized by $\alpha$.
Given $\bar{m}_{ui}$, the optimal $\omega$ is computed as:  
\begin{alignat}{2}
\omega^*(\alpha) = \mathop{\arg\min}_{\omega} L_{train}(\omega, \alpha)
\end{alignat}
The exposure parameters $\alpha$ can be seen as a special type of hyper-parameter and we update $\alpha$ automatically by minimizing the validation loss $L_{val}(\omega^*(\alpha))$ on the unbiased validation set.
%and learned automatically with guidance from the unbiased validation set.
%
% Specifically, we select the most popular positive item and negative item for each active user to form the unbiased validation set. 
% %
% The reason why the validation set can be treated as unbiased is that these items are very likely to be exposed to these users and we approximate $m_{ui}$ as 1 in the validation set. 
% %
%We update $\alpha$ automatically by minimizing the validation loss $L_{val}(\omega^*(\alpha))$.
%
Note that $L_{val}(\omega^*(\alpha))$ does not explicitly contain any $\alpha$ term since the user-item exposure $m_{ui}$ is approximated as 1 in the unbiased validation set.

Our formulation implies a bi-level optimization problem with exposure parameters $\alpha$ as the outer variable and the model parameters $\omega$ as the inner variable:
\begin{alignat}{2}
\min_{\alpha} \quad & L_{val}(\omega^*(\alpha)) & \tag{1UBO}\\
\mbox{s.t.}\quad & \omega^*(\alpha) = \mathop{\arg\min}_{\omega} L_{train}(\omega, \alpha) & \tag{2UBO}
\end{alignat}
For efficiency, we use a gradient step with the learning rate $\eta$ to approximate $\omega^*(\alpha)$ in the inner loop:
\begin{equation} \label{inner}
\begin{aligned}
    \omega^{*}(\alpha)  \approx \omega - \eta \frac{\partial{L_{train}(\omega, \alpha)}}{\partial{\omega}} 
\end{aligned}
\end{equation}
Similarly, in the outer loop, we update $\alpha$ by minimizing $\mathcal{L}_{val}(\omega^{*}(\alpha))$ via a gradient descent step with the outer loop learning rate $\eta^{'}$.
% \begin{equation} \label{outer}
% \begin{aligned}
%     \alpha^* \approx  \alpha - \eta^{'} \frac{\partial{L_{val}(\omega^*(\alpha))}}{\partial{\alpha}} 
% \end{aligned}
% \end{equation}
%Our bi-level optimization framework is summarized in Algorithm~\ref{alg:algorithm}.

\noindent\textbf{Interpretation by gradient analysis}.
By analyzing gradients, we give a natural interpretation of bi-level optimization in UBO.
In the validation set, assume $u$ likes $i_1$ and dislikes $i_2$ ($i_1$ or $i_2$ can not be $i$); $u_1$ likes $i$ and $u_2$ dislikes $i$ ($u_1$ or $u_2$ can not be $u$).
We first compute the gradient for $\tilde{R}_{ui}$ = 1(See Appendix B for details):
\begin{equation}
\begin{aligned}
\frac{\partial{L_{val}^{'}}}{\partial{\bar{m}_{ui}}} = 0
\end{aligned}
\end{equation}
This means $\bar{m}_{ui}$ will not be updated explicitly for the positive feedback.
Denote $\omega_u^T\omega_i$ as $\bar{R}_{ui}$ and then we compute the gradient for $\tilde{R}_{ui}$ = 0(See Appendix C for details):
\begin{equation}\label{analysis}
\begin{aligned}
\frac{\partial{L_{val}^{'}}}{\partial{m_{ui}}} &=\frac{\eta \sigma(\bar{R}_{ui}) \sigma(-\bar{R}_{ui})}{(1-\bar{m}_{ui}\sigma(\bar{R}_{ui}))^2}{[\omega_{i_1}^T\omega_i\sigma (-\bar{R}_{u{i_1}})}  \\& {- \omega_{i_2}^T\omega_i\sigma (\bar{R}_{u{i_2}}) +  \omega_{u_1}^T\omega_u\sigma (-\bar{R}_{{u_1}i}) }\\& {- \omega_{u_2}^T\omega_u\sigma (\bar{R}_{{u_2}i})]}
\end{aligned}
\end{equation}
%
%We interpret the first term in Eq.(\ref{analysis}) and the other three terms can be analyzed similarly.
%
For those $i$ similar to $i_1$, we know $u$ likes $i$ because $u$ likes $i_1$. 
Hence, the only explanation of $\tilde{R}_{ui} = 0$ is that $m_{ui}$ is so small that $u$ misses $i$.
The first term in Eq.(\ref{analysis}) leads to the same conclusion.
To be specific, given $i$ and $i_1$ are similar, $w_{i_1}^Tw_i$ is positive because $w_i$ and $w_{i_1}$ are in the same embedding space.
Then  we know $\frac{\eta \sigma(\bar{R}_{ui}) \sigma(-\bar{R}_{ui})}{(1-m_{ui}\sigma(\bar{R}_{ui}))^2}{w_{i_1}^Tw_i\sigma (-\bar{R}_{u{i_1}})}$ is positive, therefore this term contributes to the decrease of $\bar{m}_{ui}$ in the gradient descent step. 
Since all $\bar{m}_{ui}$ share the same distributed representation, the user-item exposure $\bar{m}_{ui}$ where $i$ is unsimiliar to $i_1$ will be updated automatically.
The other three terms in Eq~(\ref{analysis}) can be analyzed similarly.

%The problem of $L_1$ is the $\frac{1}{m_{ui}}$ term, and the correponding term of $L_2$ is $\log {m_{ui}}$. When $m_{ui}$ becomes small, $L_2$ is more stable than $L_2$.
\section{REAL-WORLD EXPERIMENTS}
\label{real}
In this section, we conduct experiments on two real-world datasets and compare several state-of-the-art methods with UBO. 
We aim to answer the following research questions:
\begin{itemize}
    \item RQ1: Does UBO outperform other methods?
    \item RQ2: Is bi-level optimization necessary in UBO?
\end{itemize}

\subsection{Experimental Setup}
%In this subsection, we introduce the datasets, comparison methods, evaluation protocols, and training details.

\subsubsection{Datasets} 
To be best of our knowledge, the Yahoo!R3\footnote{https://webscope.sandbox.yahoo.com/} dataset and the Coat\footnote{https://www.cs.cornell.edu/ schnabts/mnar} dataset are the only two public datasets that contain users' ratings for randomly selected items, and we use the two datasets to measure the true recommendation performance of UBO and the comparison methods.
See \cite{saito2020unbiased} for dataset details.
%
% To be specific, Yahoo contains approximately 15.4k users, 1k songs, and 300k five-star user-song ratings in the training set. 
% %
% Besides, the unbiased test set is collected by sampling a subset of 5,400 users and asking every one of them to rate 10 randomly selected songs. 
% %
% Coat contains approximately 290 users, 300 coats, and 6,500 five-star user-coat ratings in the training set.
% %
% Similar to Yahoo, the unbiased test set is collected by asking all users to rate  16 randomly selected coats.
%
%
%
Both datasets use the following preprocessing procedure.
Suggested by \cite{yang2018unbiased}, we treat ratings $\ge$ 4 as positive feedback and others as negative feedback.
We first select the most popular negative item and the most popular positive item for the most active 20\% users to form the validation set, which can be approximated as unbiased since the items are very likely to be exposed to the users.
%
%Following the suggestion in  \cite{ren2018learning}, we add the unbiased validation set into the training set since the model can always leverage more information from the unbiased validation set. 
%
Besides, we select $10\%$ data from the training set to form a hyper-validation set to tune hyperparameters. 
%

% \subsubsection{Datasets} 
% Yahoo!R3\footnote{https://webscope.sandbox.yahoo.com/} and Coat\footnote{https://www.cs.cornell.edu/ schnabts/mnar} datasets are the only two public datasets that contain users’ ratings for randomly selected items, and we use the two datasets to measure the true performance of recommender models.
% %

% Yahoo contains approximately 15.4k users, 1k songs, and 300k five-star user-song ratings in the training set. 
% %
% Besides, the unbiased test set is collected by sampling a subset of 5,400 users and asking every one of them to rate 10 randomly selected songs. 
% %
% Coat contains approximately 290 users, 300 coats, and 6,500 five-star user-coat ratings in the training set.
% %
% Similar to Yahoo, the unbiased test set is collected by asking all users to rate  16 randomly selected coats.
% %

% %
% Both datasets use the following preprocessing procedure.
% %
% Suggested by \cite{yang2018unbiased}, we treat ratings $\ge$ 4 as positive feedbacks and others as negative feedbacks.
% %
% We first select the most popular negative item and positive item for the most active 20\% users to form the validation set.
% %
% Then the validation set can be approximated as an unbiased validation set since it is very likely the items are exposed to the users.
% %
% Apart from the validation set, we select $10\%$ data from the training set to form a hyper-validation set to tune hyperparameters. 
% %
% Following the suggestion in  \cite{ren2018learning}, we add the validation set into the training set since the model can always leverage more information from the validation set. 

\subsubsection{Comparison methods}
We mainly compare UBO with the following methods:
\begin{itemize}
    \item \textbf{RelMF}~\cite{saito2020unbiased} adopts an unbiased estimator and uses the item popularity to approximate exposure.
    \item \textbf{ExpoMF}~\cite{liang2016modeling} introduces exposure variables to build a probabilistic model and estimates exposure via the Expectation-Maximization algorithm.
    \item \textbf{CJMF}~\cite{zhu2020unbiased} leverages different parts of the training dataset to jointly train multiple models for exposure estimation. 
    %Note that UBO and CJMF both adopt the cross-entropy loss in this paper.
    \item \textbf{BPR}~\cite{rendle2012bpr} is the most widely used algorithm for the top-N recommenders in implicit feedback.
    %\item \textbf{EJO}: This is our proposed method but it uses Joint Optimization instead of Bi-level Optimization.
    \item \textbf{UMF} uses the same exposure estimation as that in RelMF but adopts our unbiased low-variance estimator.
    
    % \item \textbf{UBO}: This is our proposed method that uses bi-level optimization to update exposure parameters simultaneously with the learning process of relevance parameters.
\end{itemize}

\subsubsection{Evaluation protocols}
Suggested by~\cite{saito2020unbiased}, we report the DCG@K (Discounted Cumulative Gain) and MAP@K (Mean Average Precision) to evaluate the ranking performance of all methods. 
We set K=1,2,3 in our experiments since the number of exposed items in the test set is small: Yahoo has 10 items and Coat has 16 items.

%\textbf{SONG: Can you explain the physical meaning of the two metrics here?}
\subsubsection{Training details}
We use Pytorch to implement UBO and optimize it with Adam.
%\cite{kingma2014adam}. 
%
We set the learning rate as $10^{-3}$, the hidden dim as $50$, the batch size as $1024$, the training epoch as $100$ for all methods on all datasets unless otherwise specified.
%
%Since CJMF uses $C$=$8$ models, we train CJMF for a total of 15 epochs for a fair comparison.
%
For other hyperparameters such as weight decay, we tune them via the performance on the hyper-validation set using the SNIPS  \cite{yang2018unbiased} estimator. 
We run every experiment five times and report the average.
Besides, we report one standard deviation in the Appendix D.

\subsection{RQ1: UBO outperforms other methods.}

\begin{table}
\centering
% \scriptsize
\caption{(RQ1)Experiment Results on Yahoo for Comparison.}
\label{table: Yahoo}
\scalebox{.85}{
\begin{tabular}{cccccccc}
\specialrule{\cmidrulewidth}{0pt}{0pt}
 Metrics & RelMF & ExpoMF & CJMF & BPR & UMF & UBO\\
\specialrule{\cmidrulewidth}{0pt}{0pt}
DCG@1 & $0.501$& $0.521$ & $0.535$ & ${0.534}$ &{$0.541$} & $\textbf{{0.552}}$\\
DCG@2 & $0.686$ & $0.724$ & {$0.742$} & ${0.740}$ &${0.731}$ & $\textbf{{0.766}}$\\
DCG@3 & $0.794$ & $0.849$  & {$0.866$} &${0.856}$ &${0.842}$ & $\textbf{{0.888}}$\\
MAP@1 & $0.501$ & $0.521$  & $0.532$ &${0.534}$ &{${0.541}$} & $\textbf{{0.552}}$\\
MAP@2 & $0.583$ & $0.610$  & $0.622$ &${0.622}$ &{${0.625}$} & $\textbf{{0.642}}$\\
MAP@3 & $0.608$ & $0.636$ & $0.646$ &${0.645}$&{$0.652$} & $\textbf{{0.664}}$\\
\specialrule{\cmidrulewidth}{0pt}{0pt} 
\end{tabular}
}
\end{table}

\begin{table}
\centering
% \scriptsize
\caption{(RQ1)Experiment Results on Coat for Comparison.}
\scalebox{.85}{
\begin{tabular}{ccccccc}
\toprule
Metrics  & RelMF & ExpoMF & CJMF & BPR & UMF & UBO \\
\midrule
DCG@1 & $0.555$ & $0.523$ & $0.504$ & ${0.568}$ &  ${0.556}$ & $\textbf{{0.573}}$\\
 DCG@2 & $0.739$ & $0.728$ & $0.729$ &{${0.770}$} & ${0.742}$ & $\textbf{{0.792}}$\\
DCG@3 & $0.887$ & $0.851$  & $0.873$ &{${0.906}$} & ${0.896}$ & $\textbf{{0.931}}$\\
MAP@1 & $0.555$ & $0.523$  & $0.504$ &{${0.568}$} & ${0.556}$ & $\textbf{{0.573}}$\\
MAP@2 & $0.622$ & $0.594$  & $0.589$ &{${0.635}$} & ${0.621}$ & $\textbf{{0.648}}$\\
MAP@3 & $0.636$ & $0.606$ & $0.603$ &{${0.650}$} &${0.639}$& $\textbf{{0.659}}$\\
\bottomrule
\end{tabular}
}
\label{table: Coat}
\vspace{-10pt}
\end{table}
%\textbf{SONG: We take what experiment to evaluate what}.
In this subsection, we aim to answer RQ1: Does UBO outperform other methods? 
Table~\ref{table: Yahoo} and Table \ref{table: Coat}  show the performances for all six methods including UBO on Yahoo and Coat respectively. 

Firstly, we observe UBO achieves the best performance among all methods on the two datasets.
This verifies the effectiveness of UBO.
Secondly, UMF outperforms RelMF in DCG@3 by about $7.4$\% in Yahoo.
This can be explained by the high gradient variance of RelMF. 
High gradient variance causes inaccurate gradient updates and thus reduces the recommendation performance.
Note that we cannot visualize the gradient variance since the gradient variance comes from the assumption of the randomness of the dataset. 
A single dataset can be seen as a single
data point and thus cannot compute its variance.
The advantage of UMF over RelMF in Coat is smaller.
The reason may be that the size of Coat is small and can not reveal the difference between UMF and RelMF.
Thirdly, UBO outperforms UMF in DCG@3 by about $4$\% in both datasets because UBO connects exposure estimation not only with the item information but also with the user information. 
%
%We also observe that BPR is a strong baseline which is consistent with our previous experience.
%
See Appendix E for time complexity discussion.

% As for the time complexity, in Yahoo, RelMF takes
% $1.35$ seconds to finish one epoch on the GeForce GTX$1650$
% platform and UBO takes $6.54$ seconds due to the bi-level optimization computing process.
% %
% Also, we have done some experiments, which apply clipping
% parameters\cite{saito2020unbiased} such as 0.01 on RelMF, CJMF, UMF and UBO. 
% %
% Results are similar to the ones without clipping parameters and the reason may be reducing variance while introducing bias.

\subsection{RQ2: Necessity of bi-level optimization}
\begin{figure}[htbp]
\centering
\includegraphics[width=0.8\columnwidth]{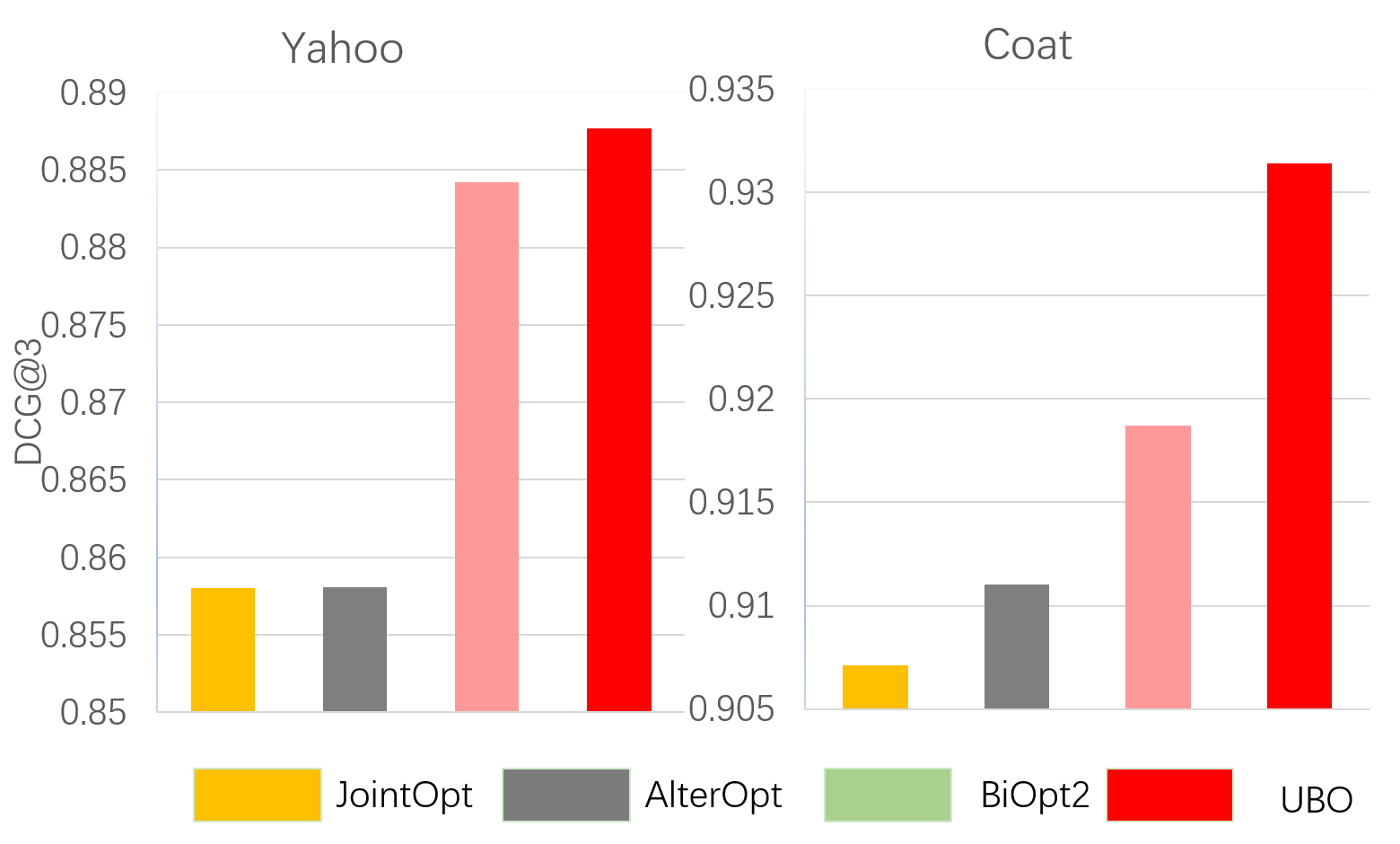}
\caption{(RQ2)Experiment Results on Yahoo and Coat. 
%
% JointOpt and AlterOpt yield similar results, and UBO outperforms both of them in DCG3 by around 3\% in both datasets.
% %
% The reason is that JointOpt and AlterOpt receive no guidance from the unbiased validation set when estimating exposure.
% %
% Similarly, BiOpt2 does not leverage the information of the unbiased validation set and thus is worse than UBO.
}
\vspace{-10pt}
\label{fig: RQ2-Yahoo-Coat}
\end{figure}

\noindent In this subsection, we aim to answer RQ2: Is bi-level optimization necessary in our method?
To better understand the necessity of bi-level optimization in UBO, we investigate two baseline strategies, where the exposure parameters and the relevance parameters are jointly optimized and alternately optimized, respectively. 
We denote the two baseline strategies as \textbf{JointOpt} and \textbf{AlterOpt} respectively. 
As we can see in  Figure \ref{fig: RQ2-Yahoo-Coat}, JointOpt and AlterOpt yield similar results, and UBO outperforms both of them in DCG3 by around 3\% in both datasets.
The reason is that JointOpt and AlterOpt
do not leverage the information of the unbiased validation set when updating the exposure parameters.
%without the guidance from the unbiased validation set.

% \begin{figure}
%     \centering
%     \subfigure{
%     \begin{minipage}[b]{0.4\textwidth}
%     \includegraphics[width=.95\textwidth]{figures/0-ml.PNG}
%     \label{fig: attention}
%     \end{minipage}
%     }
%     %\vspace{-5pt}
%     \subfigure{
%     \begin{minipage}[b]{0.4\textwidth}
%     \includegraphics[width=.95\textwidth]{figures/amazon.PNG}
%     \end{minipage}
%     \label{fig: weight}
%     }
%     \caption{(RQ3)This figure shows the trend of the PCC value between the ground-truth exposure and the learned exposure on the ML 100k and Amazon CDs datasets for ExpoMF, CJMF, JointOpt, and UBO.}
%     %\vspace{-5pt}
%     %\label{fig: RQ3-ML100k-AmazonCDs-P}
%     \label{fig: RQ3-Exposure}
% \end{figure}

Inspired by  \cite{ma2020probabilistic}, we also consider another baseline strategy of bi-level optimization. 
Instead of using the unbiased validation set, we treat every train batch as the validation set and perform bi-level optimization:
\vspace{-3pt}
\begin{alignat}{2}
\min_{\alpha} \quad & L_{train}(\omega^*(\alpha), \alpha) & \tag{1BiOpt2}\\
\mbox{s.t.}\quad & \omega^*(\alpha) = \mathop{\arg\min}_{\omega} L_{train}(\omega, \alpha) & \tag{2BiOpt2}
\end{alignat}
Note that exposure in the validation set can not be approximated as 1 anymore, so we use the estimated $\bar{m}_{ui}$ to represent the user-item exposure.
We denote this new bi-level optimization strategy as \textbf{BiOpt2}. 
BiOpt2 receives no guidance from the unbiased validation set and thus is worse than UBO.
BiOpt2 improves DCG3 over JointOpt and AlterOpt, by around 3\% in Yahoo and 1\% in coat. 
The reason may be that BiOpt2 considers the relation between $\omega$ and $\alpha$, which narrows down optimization space to a more reasonable one and thus improves training similar to \cite{ghosh2021we}.
%
%report similar results.
%where bi-level optimization on a biased validation set achieves satisfying performance.

\section{SEMI-SYNTHETIC EXPERIMENTS}
\label{semi}

%Besides the real-world dataset experiments which have verified the effectiveness of UBO
We further investigate the correctness of the estimated exposure of UBO on semi-synthetic datasets. Specifically, we aim to answer RQ3: Does UBO learn exposure correctly?

\subsection{Datasets}
%We use MovieLens (ML) 100K\footnote{https://grouplens.org/datasets/movielens/100k/}  and Amazon CDs\footnote{http://snap.stanford.edu/data/amazon/} to construct semi-synthetic datasets. 
%
% ML 100k is collected by a movie website and contains five-star movie ratings for 1683 movies by 944 users. 
% %
% Amazon has 3,749,004 five-star ratings for 486,360 items by 1,578,597 users.
%
%For Amazon, we remove users who have less than 10 interactions and items which has less than 8 interactions. 
%
%To avoid memory errors in creating the semi-synthetic dataset, we only keep the first 3,000 users and 3,000 items for Amazon. 
%
%
To answer RQ3, we need to know ground-truth exposure parameters in the dataset. 
Similar to  \cite{schnabel2016recommendations}\cite{saito2020unbiased}, we create two semi-synthetic datasets based on MovieLens (ML) 100K and Amazon CDs respectively.
See Appendix F for details.
% %
% The preprocessing procedure of semi-synthetic datasets is the same as that of Yahoo.
%

\subsection{Training and Evaluation}
 Denote $\bar{m}_{ui}$ as the estimated exposure between $u$ and $i$.
To measure the correlation between the estimated exposure $\bar{m}_{ui}$ and the true exposure $m_{ui}$, we introduce Pearson Correlation Coefficient(PCC) \cite{wright1921correlation}.
The PCC value ranges from -1 to 1. 
A value approximating to 1 means a strong positive linear relationship between the two variables, and a value approximating to -1 means a strong negative linear relationship. 
A zero value means no linear correlation between the two variables.
For every user $u$, we compute the PCC value between $\bar{m}_{ui}$ and ${m}_{ui}$ against all $M$ items.
% \begin{scriptsize}
% \begin{equation}
% pcc_{u} = \frac{M\sum_{i} m_{ui}\bar{m}_{ui} - \sum_{i} m_{ui} \sum_{i} \bar{m}_{ui}}{\sqrt{M\sum_{i} m_{ui}^2-(\sum_{i} m_{ui})^2}\sqrt{M\sum_{i} \bar{m}_{ui}^2-(\sum_{i} \bar{m}_{ui})^2}}.
% \end{equation}
% \end{scriptsize}
%
%
We report the average PCC for all users.
%
%To speed up training in Amazon, we use a batch size 8096 instead of 1024.
%
%The rest of the training and evaluation procedure is similar to that of the real-world experiments.
%

\subsection{RQ3: Does UBO learn exposure correctly?}

% \begin{figure*}[t]
%     \begin{minipage}{.9\linewidth}
%     \centering
%     \hspace{-2ex}
%     \subfigure{\label{fig: soft}
%     \includegraphics[width=0.40\textwidth]{figures/0-ml.png}}
%     \subfigure{\label{fig: lr}
%     \includegraphics[width=0.40\textwidth]{figures/amazon.png}}
% \vspace{-6pt}
%     \end{minipage}
%     \caption{%
%     This figure shows the trend of the PCC value between the ground-truth exposure and the learned exposure on the ML 100k and Amazon CDs datasets for ExpoMF, CJMF, JointOpt, and UBO.
%     }
% \vspace{-8pt}
%     \label{fig: RQ3-Exposure}
% \end{figure*}

\begin{figure}
    \centering
    \subfigure{
    \begin{minipage}[b]{0.4\textwidth}
    \includegraphics[width=1.0\textwidth]{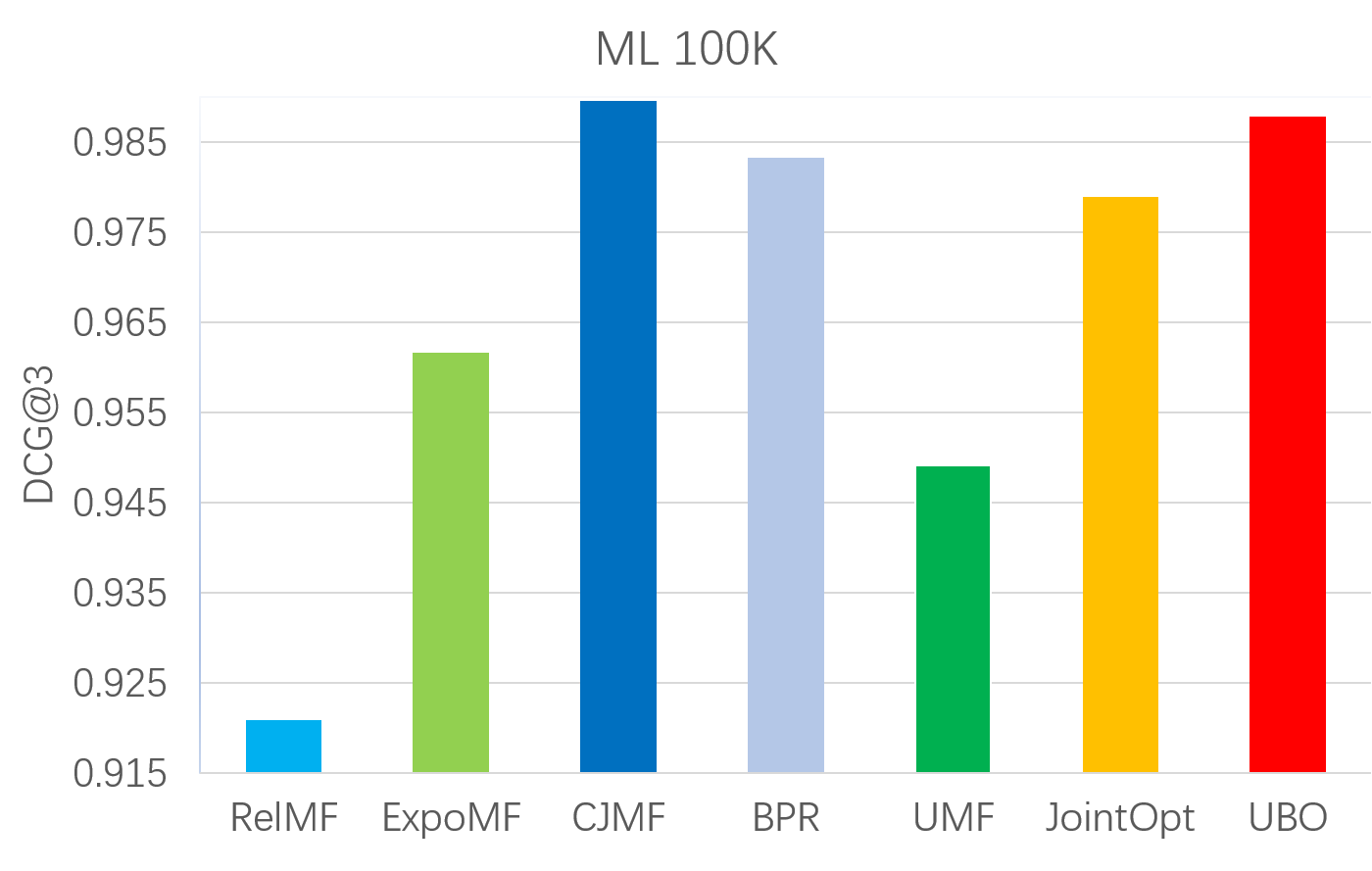}
    \label{fig: attention}
    \end{minipage}
    }
    %\vspace{-5pt}
    \subfigure{
    \begin{minipage}[b]{0.4\textwidth}
    \includegraphics[width=1.0\textwidth]{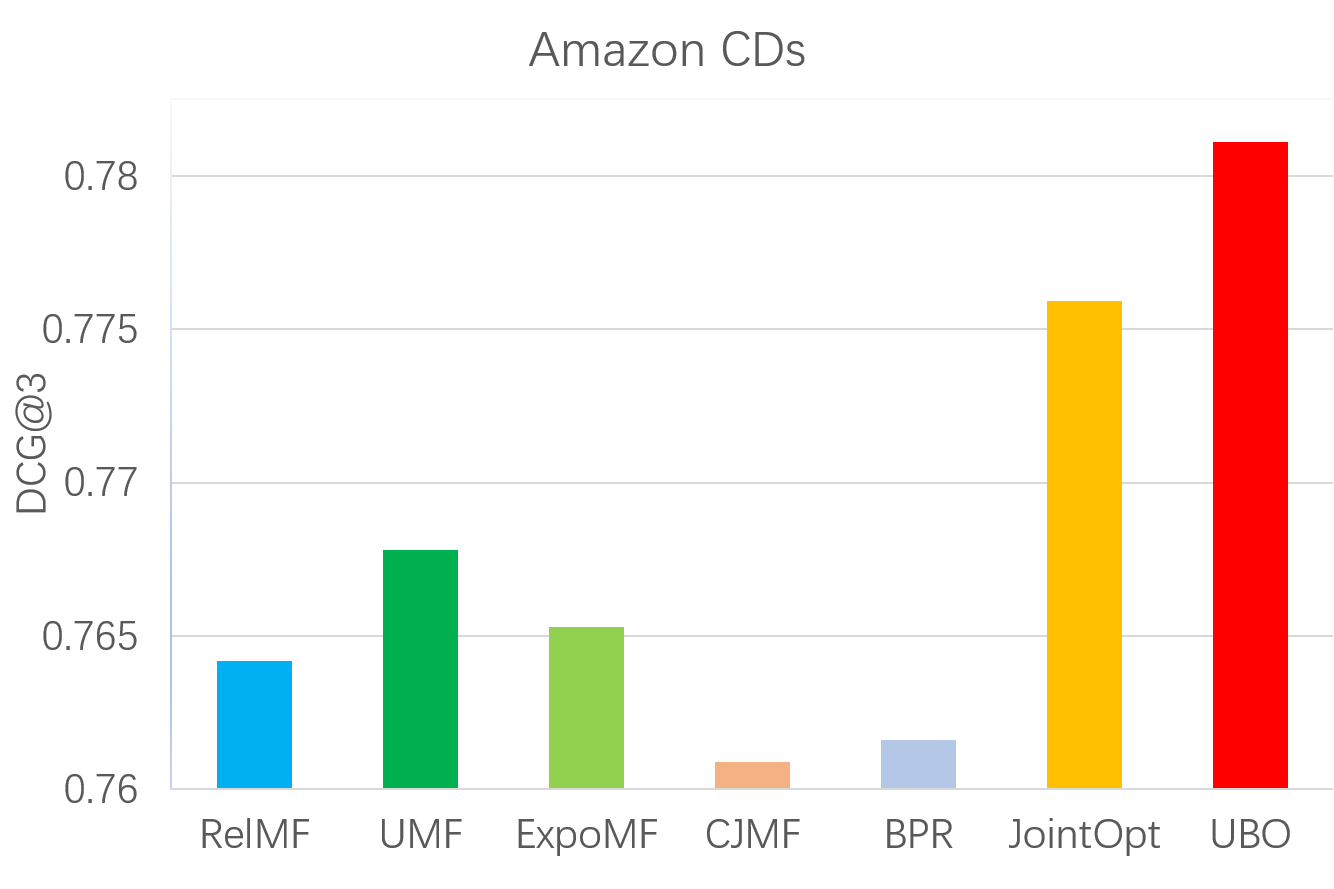}
    \end{minipage}
    \label{fig: weight}
    }
    \caption{(RQ3)Experiment Results on ML100k and Amazon CDs.}
    %\vspace{-5pt}
    \label{fig: RQ3-ML100k-AmazonCDs-P}
    %\label{fig: RQ3-Exposure}
    \vspace{-10pt}
\end{figure}

In experiments, we find the performance of JointOpt is very similar to that of AlterOpt so we only report the results of JointOpt. 
We analyze the PCC value for ExpoMF, CJMF, JointOpt, and UBO since only the four methods estimate exposure during training.
For Amazon, we find the exposure estimated by ExpoMF barely changed in the whole training process and the exposure updating frequency of ExpoMF is much lower.
To better visualize the trend for all four methods, we only plot the PCC line for the first 100 iterations and use a straight line with the mean PCC value to represent the PCC line of ExpoMF.
%
%For CJMF, we use the code in \cite{zhu2020unbiased} and average the estimated exposure on the $C=8$ models as the final estimated exposure.

\noindent\textbf{Performance}. 
%
% \begin{figure}
%     \centering
%     \subfigure{
%     \begin{minipage}[b]{0.4\textwidth}
%     \includegraphics[width=.95\textwidth]{figures/0-RQ3-ML-P.PNG}
%     \label{fig: attention}
%     \end{minipage}
%     }
%     %\vspace{-5pt}
%     \subfigure{
%     \begin{minipage}[b]{0.4\textwidth}
%     \includegraphics[width=.95\textwidth]{figures/RQ3-Amazon-P.PNG}
%     \end{minipage}
%     \label{fig: weight}
%     }
%     \caption{(RQ3)Experiment Results on ML100k and Amazon
% CDs for Comparison.}
%     %\vspace{-5pt}
%     \label{fig: RQ3-ML100k-AmazonCDs-P}
% \end{figure}
\begin{figure*}[t]
    \begin{minipage}{.9\linewidth}
    \centering
    \hspace{6ex}
    \subfigure{
    \label{fig: soft}
    \includegraphics[width=0.40\textwidth]{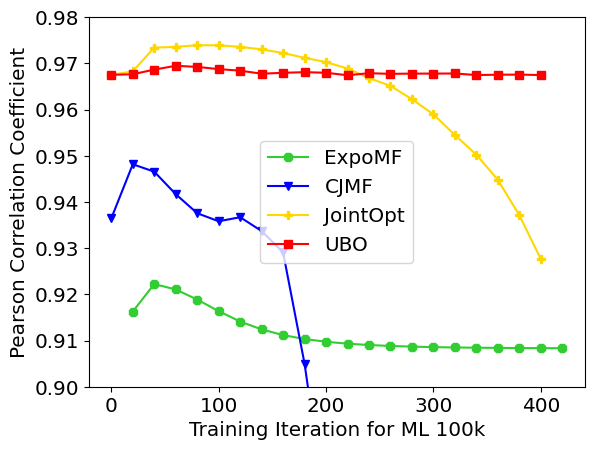}}
    \hspace{10ex}
    \subfigure{\label{fig: lr}
    \includegraphics[width=0.40\textwidth]{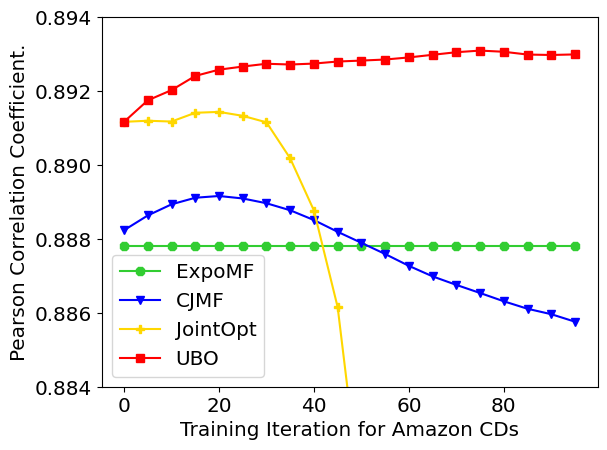}}
\vspace{-6pt}
    \end{minipage}
    \caption{%
    (RQ3)Trend of the PCC value between the ground-truth exposure and the learned exposurefor ExpoMF, CJMF, JointOpt, and UBO.
    }
\vspace{-8pt}
    \label{fig: RQ3-Exposure}
    %\label{fig: RQ3-ML100k-AmazonCDs-P}
\end{figure*}
% \begin{figure}
%     \centering
%     \subfigure{
%     \begin{minipage}[b]{0.4\textwidth}
%     \includegraphics[width=.95\textwidth]{figures/0-ml.PNG}
%     \label{fig: attention}
%     \end{minipage}
%     }
%     %\vspace{-5pt}
%     \subfigure{
%     \begin{minipage}[b]{0.4\textwidth}
%     \includegraphics[width=.95\textwidth]{figures/amazon.PNG}
%     \end{minipage}
%     \label{fig: weight}
%     }
%     \caption{(RQ3)This figure shows the trend of the PCC value between the ground-truth exposure and the learned exposure on the ML 100k and Amazon CDs datasets for ExpoMF, CJMF, JointOpt, and UBO.}
%     %\vspace{-5pt}
%     %\label{fig: RQ3-ML100k-AmazonCDs-P}
%     \label{fig: RQ3-Exposure}
% \end{figure}
%
In Figure \ref{fig: RQ3-ML100k-AmazonCDs-P} we observe that UBO is still the best performing method in Amazon CDs and outperforms other methods except for CJMF on ML 100k.
One possible explanation is that CJMF leverage $C=8$ models and one residual component simultaneously, which improves training.
UMF achieves better performances than RelMF on both datasets due to low gradient variance.

\noindent\textbf{PCC Trend Analysis}. Firstly, the PPC values in Figure \ref{fig: RQ3-Exposure} for all four methods are larger than 0.8 at early training iterations, which indicates a strong positive linear correlation.
This means all methods can estimate exposure correctly.
Besides, we observe the ExpoMF line is the lowest at the early training iterations and this is consistent with the unsatisfying results of ExpoMF in Figure \ref{fig: RQ3-ML100k-AmazonCDs-P}.
The reason may be that ExpoMF is a biased estimator ~\cite{saito2020unbiased}.
%and the estimated exposure is not accurate.

Furthermore, we compare JointOpt with UBO.
In Amazon CDs, the UBO line is higher than JointOpt all the time, in accordance with that UBO outperforms JointOpt in Figure \ref{fig: RQ3-ML100k-AmazonCDs-P}.
For ML 100k, although the PCC line of JointOpt is higher than UBO at the early training iterations, the JointOpt line experiences a gradual decrease. 
In real-world datasets, we do not have access to the PCC value which relies on the ground-truth exposure so we can not stop the training process early to get a good result of JointOpt. 
After some training iterations, the JointOpt line becomes very low in accordance with the results in Figure \ref{fig: RQ3-ML100k-AmazonCDs-P}.
The explanation may be that JointOpt updates exposure parameters and relevance parameters on the training set simultaneously, and thus experiences instability during
training.
In contrast, guided by a small unbiased validation set, UBO can enjoy a stable training process and thus estimate exposure more accurately than JointOpt.  

Last but not least, we make a comparison between CJMF and UBO.
In Amazon, the low PCC line of CJMF in Figure~\ref{fig: RQ3-Exposure} corresponds to the unsatisfying results in Figure~\ref{fig: RQ3-ML100k-AmazonCDs-P}.
Yet, on ML 100k, CJMF outperforms UBO in terms of performance in Figure~\ref{fig: RQ3-ML100k-AmazonCDs-P} while the PCC line of CJMF is lower than that of UBO in Figure~\ref{fig: RQ3-Exposure}.
The reason may be CJMF leverages an extra residual component to improve training, which is not included in the exposure estimation process in Figure~\ref{fig: RQ3-Exposure}.

\section{RELATED WORK}
\label{related}
%
%Our work is inspired by two lines of research in recommender systems: 1) debiasing and 2) bi-level optimization.

%\subsection{Debiasing}

Many important work~\cite{steck2010training} \cite{steck2013evaluation}\cite{hernandez2014probabilistic}\cite{wang2018deconfounded}\cite{wang2019doubly} \cite{joachims2016counterfactual} \cite{wang2020causal} \cite{liang2016causal} \cite{bonner2018causal}\cite{schnabel2016recommendations} have studied the bias in the explicit rating data.
%in the recommender systems.
%
For example, as the user can choose which items to rate freely, the observed ratings cannot serve as a representative sample of all ratings.
Thus the biased rating data leads to challenges for both recommendation evaluation and training.
To correct this bias, many methods~\cite{wang2018deconfounded}\cite{wang2019doubly}\cite{joachims2016counterfactual}\cite{wang2020causal}\cite{liang2016causal} use causal inference to learn from biased data and achieve better recommendation performances.

Compared with explicit feedback, implicit feedback is much easier to collect and thus plays a more important role, which renders debiasing in implicit feedback an important topic.
%
% debias in the evaluation of recommender system,
\cite{yang2018unbiased} develop an unbiased offline evaluator which significantly reduces the bias toward popular items
%first show the widely used Average-Over-All (AOA) evaluator is biased toward popular items and then develop an unbiased offline evaluator which significantly reduces the bias.
%
To debias in model training, \cite{hu2008collaborative} \cite{devooght2015dynamic} adopt a heuristic-based strategy, where unobserved interactions are assigned with a lower weight.
Furthermore, \cite{pan2008one,pan2009mind} associate the weight with the user's activity and~\cite{he2016fast}\cite{yu2017selection} specify the weight with the item popularity.
\cite{gupta2021correcting} propose to leverage known exposure probabilities to mitigate exposure bias for link prediction.
From a casual perspective, \cite{liang2016modeling} 
directly incorporate exposure into collaborative filtering and build a probabilistic model.
%
%Yet, \cite{saito2020unbiased} find \cite{liang2016causal} biased towards popular items and yield unsatisfied results for rare items.
%
Based on the IPS technique, \cite{saito2020unbiased} propose an unbiased estimator with the item popularity as exposure estimation.
%by leveraging the IPS technique.
%
For better exposure estimation, \cite{zhu2020unbiased} propose a combinatorial joint learning framework to solve the estimation-training overlap problem.
The estimated exposure can still be biased since it only leverages biased training data. 
Besides, we find the unbiased estimator in~\cite{saito2020unbiased,zhu2020unbiased} suffers from the high gradient variance problem. 
In this paper, we propose an unbiased estimator with low variance from a probabilistic view.
\cite{chen2021autodebias} leverages another set of data to debias data by solving the bi-level optimization problem.
The main differences between \cite{chen2021autodebias} and UBO are a) UBO has a low-variance unbiased estimator while \cite{chen2021autodebias} does not and b) \cite{chen2021autodebias} requires an unbiased set in advance, while UBO forms the unbiased set from the training set, which means we can not directly compare the two methods.

\section{CONCLUSION}
\label{conclusion}
To bridge the gap between the implicit feedback and the user-item relevance, existing approaches explicitly model the user-item exposure while the proposed unbiased estimators suffer from high gradient variance.
In this paper, we propose a low-variance unbiased estimator from a probabilistic view and this estimator effectively bounds the gradient variance.
Besides, we connect exposure estimation with both user and item information and then collect an unbiased set to guide exposure estimation.
By leveraging the unbiased set, we update exposure parameters and relevance parameters simultaneously via bi-level optimization.
%
%We conduct experiments on both real-world datasets and semi-synthetic datasets to verify the effectiveness of UBO.
Experiments on real-world datasets and semi-synthetic datasets verify the effectiveness of UBO.

%bibliographystyle{named}
%\bibliography{nbo}

\bibliographystyle{named}
\bibliography{ijcai22}

\begin{thebibliography}{}

\bibitem[\protect\citeauthoryear{Bonner and Vasile}{2018}]{bonner2018causal}
Stephen Bonner and Flavian Vasile.
\newblock Causal embeddings for recommendation.
\newblock In {\em In Proc. of the 12th ACM Conference on RecSys}, 2018.

\bibitem[\protect\citeauthoryear{Chen \bgroup \em et al.\egroup
  }{2021}]{chen2021autodebias}
Jiawei Chen, Hande Dong, Yang Qiu, Xiangnan He, Xin Xin, Liang Chen, Guli Lin,
  and Keping Yang.
\newblock Autodebias: Learning to debias for recommendation.
\newblock {\em arXiv preprint arXiv:2105.04170}, 2021.

\bibitem[\protect\citeauthoryear{Colson \bgroup \em et al.\egroup
  }{2007}]{colson2007overview}
Beno{\^\i}t Colson, Patrice Marcotte, and Gilles Savard.
\newblock An overview of bilevel optimization.
\newblock {\em Annals of operations research}, 2007.

\bibitem[\protect\citeauthoryear{Devooght \bgroup \em et al.\egroup
  }{2015}]{devooght2015dynamic}
Robin Devooght, Nicolas Kourtellis, and Amin Mantrach.
\newblock Dynamic matrix factorization with priors on unknown values.
\newblock In {\em Proc. of KDD}, 2015.

\bibitem[\protect\citeauthoryear{Ghosh and Lan}{2021}]{ghosh2021we}
Aritra Ghosh and Andrew Lan.
\newblock Do we really need gold samples for sample weighting under label
  noise?
\newblock In {\em Proceedings of the IEEE/CVF Winter Conference on Applications
  of Computer Vision}, 2021.

\bibitem[\protect\citeauthoryear{Gupta \bgroup \em et al.\egroup
  }{2021}]{gupta2021correcting}
Shantanu Gupta, Hao Wang, Zachary~C Lipton, and Yuyang Wang.
\newblock Correcting exposure bias for link recommendation.
\newblock In {\em Proc. of ICML}, 2021.

\bibitem[\protect\citeauthoryear{He \bgroup \em et al.\egroup
  }{2016}]{he2016fast}
Xiangnan He, Hanwang Zhang, Min-Yen Kan, and Tat-Seng Chua.
\newblock Fast matrix factorization for online recommendation with implicit
  feedback.
\newblock In {\em Proc. of SIGIR}, pages 549--558, 2016.

\bibitem[\protect\citeauthoryear{He \bgroup \em et al.\egroup
  }{2017}]{he2017neural}
Xiangnan He, Lizi Liao, Hanwang Zhang, Liqiang Nie, Xia Hu, and Tat-Seng Chua.
\newblock Neural collaborative filtering.
\newblock In {\em Proc. of WWW}, 2017.

\bibitem[\protect\citeauthoryear{Hern{\'a}ndez-Lobato \bgroup \em et al.\egroup
  }{2014}]{hernandez2014probabilistic}
Jos{\'e}~Miguel Hern{\'a}ndez-Lobato, Neil Houlsby, and Zoubin Ghahramani.
\newblock Probabilistic matrix factorization with non-random missing data.
\newblock In {\em Proc. of ICML}, 2014.

\bibitem[\protect\citeauthoryear{Hu \bgroup \em et al.\egroup
  }{2008}]{hu2008collaborative}
Yifan Hu, Yehuda Koren, and Chris Volinsky.
\newblock Collaborative filtering for implicit feedback datasets.
\newblock In {\em Proc. of ICDM}, pages 263--272. Ieee, 2008.

\bibitem[\protect\citeauthoryear{Joachims and
  Swaminathan}{2016}]{joachims2016counterfactual}
Thorsten Joachims and Adith Swaminathan.
\newblock Counterfactual evaluation and learning for search, recommendation and
  ad placement.
\newblock In {\em Proc. of SIGIR}, 2016.

\bibitem[\protect\citeauthoryear{Koren \bgroup \em et al.\egroup
  }{2009}]{koren2009matrix}
Yehuda Koren, Robert Bell, and Chris Volinsky.
\newblock Matrix factorization techniques for recommender systems.
\newblock {\em Computer}, 2009.

\bibitem[\protect\citeauthoryear{Liang \bgroup \em et al.\egroup
  }{2016a}]{liang2016causal}
Dawen Liang, Laurent Charlin, and David~M Blei.
\newblock Causal inference for recommendation.
\newblock In {\em Causation: Foundation to Application, Workshop at UAI. AUAI},
  2016.

\bibitem[\protect\citeauthoryear{Liang \bgroup \em et al.\egroup
  }{2016b}]{liang2016modeling}
Dawen Liang, Laurent Charlin, James McInerney, and David~M Blei.
\newblock Modeling user exposure in recommendation.
\newblock In {\em Proc. of WWW}, 2016.

\bibitem[\protect\citeauthoryear{Ma \bgroup \em et al.\egroup
  }{2020}]{ma2020probabilistic}
Chen Ma, Liheng Ma, Yingxue Zhang, Ruiming Tang, Xue Liu, and Mark Coates.
\newblock Probabilistic metric learning with adaptive margin for top-k
  recommendation.
\newblock In {\em Proc. of KDD}, 2020.

\bibitem[\protect\citeauthoryear{Pan and Scholz}{2009}]{pan2009mind}
Rong Pan and Martin Scholz.
\newblock Mind the gaps: weighting the unknown in large-scale one-class
  collaborative filtering.
\newblock In {\em Proc. of KDD}, 2009.

\bibitem[\protect\citeauthoryear{Pan \bgroup \em et al.\egroup
  }{2008}]{pan2008one}
Rong Pan, Yunhong Zhou, Bin Cao, Nathan~N Liu, Rajan Lukose, Martin Scholz, and
  Qiang Yang.
\newblock One-class collaborative filtering.
\newblock In {\em Proc. of ICDM}, 2008.

\bibitem[\protect\citeauthoryear{Rendle \bgroup \em et al.\egroup
  }{2012}]{rendle2012bpr}
Steffen Rendle, Christoph Freudenthaler, Zeno Gantner, and Lars Schmidt-Thieme.
\newblock Bpr: Bayesian personalized ranking from implicit feedback.
\newblock {\em arXiv preprint arXiv:1205.2618}, 2012.

\bibitem[\protect\citeauthoryear{Saito \bgroup \em et al.\egroup
  }{2020}]{saito2020unbiased}
Yuta Saito, Suguru Yaginuma, Yuta Nishino, Hayato Sakata, and Kazuhide Nakata.
\newblock Unbiased recommender learning from missing-not-at-random implicit
  feedback.
\newblock In {\em Proc. of WSDM}, 2020.

\bibitem[\protect\citeauthoryear{Schnabel \bgroup \em et al.\egroup
  }{2016}]{schnabel2016recommendations}
Tobias Schnabel, Adith Swaminathan, Ashudeep Singh, Navin Chandak, and Thorsten
  Joachims.
\newblock Recommendations as treatments: Debiasing learning and evaluation.
\newblock {\em arXiv preprint arXiv:1602.05352}, 2016.

\bibitem[\protect\citeauthoryear{Steck}{2010}]{steck2010training}
Harald Steck.
\newblock Training and testing of recommender systems on data missing not at
  random.
\newblock In {\em Proc. of KDD}, 2010.

\bibitem[\protect\citeauthoryear{Steck}{2013}]{steck2013evaluation}
Harald Steck.
\newblock Evaluation of recommendations: rating-prediction and ranking.
\newblock In {\em Proc. of RecSys}, 2013.

\bibitem[\protect\citeauthoryear{Wang \bgroup \em et al.\egroup
  }{2018}]{wang2018deconfounded}
Yixin Wang, Dawen Liang, Laurent Charlin, and David~M Blei.
\newblock The deconfounded recommender: A causal inference approach to
  recommendation.
\newblock {\em arXiv preprint arXiv:1808.06581}, 2018.

\bibitem[\protect\citeauthoryear{Wang \bgroup \em et al.\egroup
  }{2019a}]{wang2019neural}
Xiang Wang, Xiangnan He, Meng Wang, Fuli Feng, and Tat-Seng Chua.
\newblock Neural graph collaborative filtering.
\newblock In {\em Proc. of SIGIR}, 2019.

\bibitem[\protect\citeauthoryear{Wang \bgroup \em et al.\egroup
  }{2019b}]{wang2019doubly}
Xiaojie Wang, Rui Zhang, Yu~Sun, and Jianzhong Qi.
\newblock Doubly robust joint learning for recommendation on data missing not
  at random.
\newblock In {\em Proc. of ICML}, 2019.

\bibitem[\protect\citeauthoryear{Wang \bgroup \em et al.\egroup
  }{2020}]{wang2020causal}
Yixin Wang, Dawen Liang, Laurent Charlin, and David~M Blei.
\newblock Causal inference for recommender systems.
\newblock In {\em Fourteenth ACM Conference on Recommender Systems}, pages
  426--431, 2020.

\bibitem[\protect\citeauthoryear{Wright}{1921}]{wright1921correlation}
Sewall Wright.
\newblock Correlation and causation.
\newblock {\em J. agric. Res.}, 1921.

\bibitem[\protect\citeauthoryear{Yang \bgroup \em et al.\egroup
  }{2018}]{yang2018unbiased}
Longqi Yang, Yin Cui, Yuan Xuan, Chenyang Wang, Serge Belongie, and Deborah
  Estrin.
\newblock Unbiased offline recommender evaluation for missing-not-at-random
  implicit feedback.
\newblock In {\em Proc. of RecSys}, 2018.

\bibitem[\protect\citeauthoryear{Yu \bgroup \em et al.\egroup
  }{2017}]{yu2017selection}
Hsiang-Fu Yu, Mikhail Bilenko, and Chih-Jen Lin.
\newblock Selection of negative samples for one-class matrix factorization.
\newblock In {\em Proc. of SDM}, 2017.

\bibitem[\protect\citeauthoryear{Zhu \bgroup \em et al.\egroup
  }{2020}]{zhu2020unbiased}
Ziwei Zhu, Yun He, Yin Zhang, and James Caverlee.
\newblock Unbiased implicit recommendation and propensity estimation via
  combinational joint learning.
\newblock In {\em Fourteenth ACM Conference on Recommender Systems}, 2020.

\end{thebibliography}

\end{document}

\renewcommand\thesection{\Alph{section}} 
\title{IJCAI 2041 Appendix}
\maketitle

\section{Unbiased Estimator}
Our estimator is defined as:
\begin{equation}\label{estimator21}
\begin{aligned}
\mathcal{L}_2(\omega) & = -\sum_{(u,i)\in \mathcal{D}}
{\tilde{R}_{ui}\log(\bar{m}_{ui}p_{ui}) \\& + (1-\tilde{R}_{ui})\log (1-\bar{m}_{ui}p_{ui})}
\end{aligned}
\end{equation}
The expectation of $\mathcal{L}_2(\omega)$ is calculated as:
\begin{equation}\label{estimator22}
\begin{aligned}
E(\mathcal{L}_2(\omega)) & = -\sum_{(u,i)\in \mathcal{D}}{m_{ui}\gamma_{ui}\log
(\bar{m}_{ui}p_{ui}) \\& + (1-m_{ui}\gamma_{ui})\log (1-\bar{m}_{ui}p_{ui})}
\end{aligned}
\end{equation}

Given an accurate exposure estimation $\bar{m}_{ui} = m_{ui}$, we know the optimal solution which minimizes $E(\mathcal{L}_2(\omega))$ is $p_{ui}=\gamma_{ui}$, which proves our estimator unbiased.

\section{Zero Gradient}
For analysis convenience, we treat every $\bar{m}_{ui}$ as a learnable parameter instead of using MLP to parameterize $\bar{m}_{ui}$.
Recall our estimator is defined as:
\begin{equation}\label{estimator1}
\begin{aligned}
\mathcal{L}_2(\omega) & = -\sum_{(u,i)\in \mathcal{D}}{\tilde{R}_{ui}\log
(\bar{m}_{ui}p_{ui}) \\& + (1-\tilde{R}_{ui})\log (1-\bar{m}_{ui}p_{ui})}
\end{aligned}
\end{equation}
For $\tilde{R}_{ui}$ = 1, we can write the loss function term related to $u$ and $i$ as:
\begin{equation}
\label{eq: zero}
\begin{aligned}
\mathcal{L}_{ui} &= -\log (\bar{m}_{ui}p_{ui})\\
&= -\log (\bar{m}_{ui}) -\log(p_{ui})
%\frac{\partial{L_{train}}}{\partial{m_{ui}}} = 0
\end{aligned}
\end{equation}
Note that the updated $\omega_u$ and $\omega_i$ do not contain any $\bar{m}_{ui}$ term since  the $-\log (\bar{m}_{ui})$ term in Eq~(\ref{eq: zero}) is a constant in the inner loop update.
% the $-\log (\bar{m}_{ui})$ term is not related to the updatatated $\omega_u$ and $\omega_i$.
%
As a result, $\mathcal{L}_{val}$ does not contain any $\bar{m}_{ui}$ term, which means:
\begin{equation}
\begin{aligned}
\frac{\partial{L_{val}}}{\partial{\bar{m}_{ui}}} = 0
\end{aligned}
\end{equation}
%
%This means $m_{ui}$ will not be updated explicitly for positive feedback.

\section{Gradient Analysis}
Note that the training loss function with regard to $\bar{m}_{ui}$ for $\tilde{R}_{ui}$ = $0$ can be written as:
\begin{equation}
    \begin{aligned}
    \mathcal{L}_{ui} = -\log(1-\bar{m}_{ui}\sigma(\bar{R}_{ui}))
    \end{aligned}
\end{equation}
where $\bar{R}_{ui} = \omega_u^T\omega_i$.
After a gradient descent step, the user embedding is updated as:
\begin{equation}
    \begin{aligned}
    \omega_{u}(\bar{m}_{ui}) =& \omega_u - \eta \frac{\partial{L_{ui}}}{\partial{\omega_u}}\\
    =& \omega_u - \eta \frac{\partial{L_{ui}}}{\partial{\sigma(\bar{R}_{ui})}}\frac{\partial{\sigma(\bar{R}_{ui})}}{\partial{\omega_u}} \\
    =& \omega_u - \eta \frac{\bar{m}_{ui}\sigma(\bar{R}_{ui})\sigma(-\bar{R}_{ui})\omega_i}{1-\bar{m}_{ui}\sigma(\bar{R}_{ui})}\\
    \end{aligned}
\end{equation}
Similarly, the item embedding is updated as:
\begin{equation}
    \begin{aligned}
    \omega_{i}(\bar{m}_{ui}) =& \omega_i - \eta \frac{\bar{m}_{ui}\sigma(\bar{R}_{ui})\sigma(-\bar{R}_{ui})\omega_u}{1-\bar{m}_{ui}\sigma(\bar{R}_{ui})} \\
    \end{aligned}
\end{equation}
Recall that in the validation set, assume $u$ likes $i_1$ and dislikes $i_2$ ($i_1$ or $i_2$ can not be $i$); $u_1$ likes $i$ and $u_2$ dislikes $i$ ($u_1$ or $u_2$ can not be $u$). 
We write the loss on the unbiased validation set regard to the related users($u_1$ and $u_2$) and the related items($i_1$ and $i_2$) as:
\begin{equation}
    \begin{aligned}
    \mathcal{L}_{val}^{'} = -\log(\sigma(\bar{R}_{ui_1})) - \log(1-\sigma(\bar{R}_{ui_2})) \\- \log(\sigma(\bar{R}_{u_1i}))- \log(1-\sigma(\bar{R}_{u_2i}))
    \end{aligned}
\end{equation}
According to the chain rule, we can write the gradient as:
\begin{equation}
    \begin{aligned}
    \frac{\partial{L_{val}}^{'}}{\partial{\bar{m}_{ui}}} =  \frac{\partial{L_{val}^{'}}}{\partial{\omega_u}}\frac{\partial{\omega_u}}{\partial{\bar{m}_{ui}}} +  \frac{\partial{L_{val}^{'}}}{\partial{\omega_i}}\frac{\partial{\omega_i}}{\partial{\bar{m}_{ui}}}
    \end{aligned}
\end{equation}
%We unroll the above four terms respectively:
We unroll the $\frac{\partial{L_{val}^{'}}}{\partial{\omega_u^T}}$ term as:
\begin{equation}
\begin{aligned}
       \frac{\partial{L_{val}^{'}}}{\partial{\omega_u}} =& -\frac{\sigma(\bar{R}_{ui_1})\sigma(-\bar{R}_{ui_1})}{\sigma(\bar{R}_{ui_1})}\omega_{i_1}^T \\ -&\frac{-\sigma(\bar{R}_{ui_2})\sigma(-\bar{R}_{ui_2})}{1-\sigma(\bar{R}_{ui_2})}\omega_{i_2}^T \\
       =& -\sigma(-\bar{R}_{ui_1})\omega_{i_1}^T + \sigma (\bar{R}_{ui_2})\omega_{i_2}^T
\end{aligned}
\end{equation}
Similarly, the $\frac{\partial{L_{val}^{'}}}{\partial{\omega_i}}$ term is unrolled as:
\begin{equation}
\begin{aligned}
       \frac{\partial{L_{val}^{'}}}{\partial{\omega_i}} =& -\frac{\sigma(\bar{R}_{u_1i})\sigma(-\bar{R}_{u_1i})}{\sigma(\bar{R}_{u_1i})}\omega_{u_1}^T \\ -&\frac{-\sigma(\bar{R}_{u_2i})\sigma(-\bar{R}_{u_2i})}{1-\sigma(\bar{R}_{u_2i})}\omega_{u_2}^T \\
       =& -\sigma(-\bar{R}_{u_1i})\omega_{u_1}^T + \sigma (\bar{R}_{u_2i})\omega_{u_2}^T
\end{aligned}
\end{equation}
We unroll the $\frac{\partial{\omega_u(\bar{m}_{ui})}}{\partial{\bar{m}_{ui}}}$ term as:
\begin{equation}
    \begin{aligned}
    \frac{\partial{\omega_u(\bar{m}_{ui})}}{\partial{\bar{m}_{ui}}}=&
     -\eta \sigma(\bar{R}_{ui}) \sigma(-\bar{R}_{ui}) \omega_i 
     \frac{\partial{\frac{\bar{m}_{ui}}{1-\bar{m}_{ui}\sigma(\bar{R}_{ui})}}}{\partial{\bar{m}_{ui}}} \\
    =& -\frac{\eta \sigma(\bar{R}_{ui})\sigma(-\bar{R}_{ui})\omega_i}{(1-\bar{m}_{ui}\sigma(\bar{R}_{ui}))^2}
    \end{aligned}
\end{equation}
Similarly, the $\frac{\partial{\omega_i(\bar{m}_{ui})}}{\partial{\bar{m}_{ui}}}$ term is unrolled as:
\begin{equation}
    \begin{aligned}
    \frac{\partial{\omega_i(\bar{m}_{ui})}}{\partial{\bar{m}_{ui}}}=&
     -\eta \sigma(\bar{R}_{ui}) \sigma(-\bar{R}_{ui}) \omega_u 
     \frac{\partial{\frac{\bar{m}_{ui}}{1-\bar{m}_{ui}\sigma(\bar{R}_{ui})}}}{\partial{\bar{m}_{ui}}} \\
    =& -\frac{\eta \sigma(\bar{R}_{ui})\sigma(-\bar{R}_{ui})\omega_u}{(1-\bar{m}_{ui}\sigma(\bar{R}_{ui}))^2}
    \end{aligned}
\end{equation}
In the end, we write $\frac{\partial{L_{val}^{'}}}{\partial{\bar{m}_{ui}}}$ as:
\begin{equation}
\begin{aligned}
\frac{\partial{L_{val}^{'}}}{\partial{\bar{m}_{ui}}} &= \frac{\partial{L_{val}^{'}}}{\partial{\omega_u}} 
\frac{\partial{\omega_u(\bar{m}_{ui})}}{\partial{\bar{m}_{ui}}} + \frac{\partial{L_{val}^{'}}}{\partial{\omega_i}} 
\frac{\partial{\omega_i(\bar{m}_{ui})}}{\partial{\bar{m}_{ui}}} \\
&= [\sigma(-\bar{R}_{ui_1})\omega_{i_1}^T - \sigma(-\bar{R}_{ui_2})\omega_{i_2}^T]\frac{\eta \sigma(\bar{R}_{ui}) \sigma(-\bar{R}_{ui})}{(1-\bar{m}_{ui}\sigma(\bar{R}_{ui}))^2}\omega_i \\
&+ [\sigma(-\bar{R}_{u_1i})\omega_{u_1}^T - \sigma(-\bar{R}_{u_2i})\omega_{u_2}^T]\frac{\eta \sigma(\bar{R}_{ui}) \sigma(-\bar{R}_{ui})}{(1-\bar{m}_{ui}\sigma(\bar{R}_{ui}))^2}\omega_u \\
&= \frac{\eta \sigma(\bar{R}_{ui}) \sigma(-\bar{R}_{ui})}{(1-\bar{m}_{ui}\sigma(\bar{R}_{ui}))^2}{[\omega_{i_1}^T\omega_i\sigma (-\bar{R}_{u{i_1}})  \\& - \omega_{i_2}^T\omega_i\sigma (\bar{R}_{u{i_2}}) +  \omega_{u_1}^T\omega_u\sigma (-\bar{R}_{{u_1}i}) \\& - \omega_{u_2}^T\omega_u\sigma (\bar{R}_{{u_2}i})]}\\
\end{aligned}
\end{equation}

\section{Standard Deviation}
\begin{table*}
\centering
% \scriptsize
\caption{(RQ1)Experiment Results on Yahoo for Comparison.}
\scalebox{.95}{
\begin{tabular}{cccccccc}
\toprule
 Metrics & RelMF & ExpoMF & CJMF & BPR & UMF & UBO\\
\midrule
DCG@1 & $0.501 \pm 0.004$& $0.521 \pm 0.007$ & $0.535 \pm 0.003$ & $\bm{0.534 \pm 0.007}$ &\uline{$0.541 \pm 0.009$} & $\textbf{\bm{0.552}} \pm \textbf{0.004}$\\
DCG@2 & $0.686 \pm 0.002$ & $0.724 \pm 0.011$ & \uline{$0.742 \pm 0.002$} & $\bm{0.740 \pm 0.007}$ &$\bm{0.731} \pm 0.010$ & $\textbf{\bm{0.766}} \pm \textbf{0.004}$\\
DCG@3 & $0.794 \pm 0.007$ & $0.849 \pm 0.008$  & \uline{$0.866 \pm 0.001$} &$\bm{0.856 \pm 0.005}$ &$\bm{0.842} \pm 0.014$ & $\textbf{\bm{0.888}} \pm \textbf{0.002}$\\
MAP@1 & $0.501 \pm 0.004$ & $0.521 \pm 0.007$  & $0.532 \pm 0.007$ &$\bm{0.534 \pm 0.009}$ &\uline{$\bm{0.541} \pm 0.009$} & $\textbf{\bm{0.552}}\pm \textbf{0.004}$\\
MAP@2 & $0.583 \pm 0.004$ & $0.610 \pm 0.009$  & $0.622 \pm 0.001$ &$\bm{0.622 \pm 0.007}$ &\uline{$\bm{0.625} \pm 0.008$} & $\textbf{\bm{0.642}}\pm \textbf{0.003}$\\
MAP@3 & $0.608 \pm 0.005$ & $0.636 \pm 0.007$ & $0.646 \pm 0.001$ &$\bm{0.645 \pm 0.006}$&\uline{$0.652 \pm 0.008$} & $\textbf{\bm{0.664}} \pm \textbf{0.003}$\\

\bottomrule
\end{tabular}
}
\label{table: Yahoo}
\end{table*}

\begin{table*}
\centering
% \scriptsize
\caption{(RQ1)Experiment Results on Coat for Comparison.}
\scalebox{.95}{
\begin{tabular}{ccccccc}
\toprule
Metrics  & RelMF & ExpoMF & CJMF & BPR & UMF & UBO \\
\midrule
DCG@1 & $0.555 \pm 0.013 $& $0.523 \pm 0.033$ & $0.504 \pm 0.022$ &\uline{$\bm{0.568} \pm 0.015$} &  $\bm{0.556} \pm 0.013$ &  $\textbf{\bm{0.573}} \pm \textbf{0.015}$\\
DCG@2 & $0.739 \pm 0.003$ & $0.728\pm 0.020$ & $0.729 \pm 0.023$ &\uline{$\bm{0.770} \pm 0.015$} & $\bm{0.742} \pm 0.015$ & $\textbf{\bm{0.792}} \pm \textbf{0.014}$\\
DCG@3 & $0.887 \pm 0.013$ & $0.851\pm 0.021$  & $0.873 \pm 0.016$ &\uline{$\bm{0.906} \pm 0.011$} & $\bm{0.896} \pm 0.018$ & $\textbf{\bm{0.931}} \pm \textbf{0.004}$\\
MAP@1 & $0.555\pm 0.013$ & $0.523\pm 0.033$  & $0.504 \pm 0.022$ &\uline{$\bm{0.568}\pm 0.015$} & $\bm{0.556} \pm 0.013$ & $\textbf{\bm{0.573}} \pm \textbf{0.015}$\\
MAP@2 & $0.622\pm 0.008$ & $0.594\pm 0.024$  & $0.589 \pm 0.017$ &\uline{$\bm{0.635}\pm 0.010$} & $\bm{0.621} \pm 0.015$ & $\textbf{\bm{0.648}} \pm \textbf{0.015}$\\
MAP@3 & $0.636\pm 0.008$ & $0.606\pm 0.024$ & $0.603 \pm 0.013$ &\uline{$\bm{0.650}\pm 0.008$} &$\bm{0.639} \pm 0.015$& $\textbf{\bm{0.659}} \pm \textbf{0.013}$\\
\bottomrule
\end{tabular}
}
\label{table: Coat}
\end{table*}
We report the standard devivation in Table~\ref{table: Yahoo} and Table~\ref{table: Coat}
\section{Time Complexity}
As for the time complexity, in the Yahoo dataset, RelMF takes
$1.35$ seconds to finish one epoch on the GeForce GTX$1650$
platform and UBO takes $6.54$ seconds due to the bi-level optimization computing process.

\section{Datasets}
We use the MovieLens (ML) 100K\footnote{https://grouplens.org/datasets/movielens/100k/}  and Amazon CDs\footnote{http://snap.stanford.edu/data/amazon/} datasets to construct semi-synthetic datasets. 
%
% The ML 100K dataset is collected by a movie website and contains five-star movie ratings for 1683 movies by 944 users. 
% %
% The Amazon CDs dataset has 3,749,004 five-star ratings for 486,360 items by 1,578,597 users.
%
For the Amazon CDs dataset, we remove users who have less than 10 interactions and items which has less than 8 interactions. 
To avoid memory errors in creating the semi-synthetic dataset, we only keep the first 3,000 users and 3,000 items for the Amazon CDs dataset. 
% \bibliographystyle{plain}
% \bibliography{aaai22}